\documentclass[twocolumn]{aastex62}
\usepackage[utf8]{inputenc}
\usepackage{float}
\usepackage{graphicx}
\usepackage{spverbatim}

\begin{document}

\title{The Metallicity Gradient and Complex Formation History of the Outermost Halo of the Milky Way}
\author[0000-0002-3567-1897]{Sarah E. Dietz}
\affiliation{Department of Physics, University of Notre Dame, Notre Dame, IN 46556, USA}
\affiliation{Joint Institute for Nuclear Astrophysics—Center for the Evolution of the Elements (JINA-CEE), USA}

\correspondingauthor{Sarah E. Dietz}
\email{sdietz@nd.edu}

\author[0000-0002-4168-239X]{Jinmi Yoon}
\affiliation{Department of Physics, University of Notre Dame, Notre Dame, IN 46556, USA}
\affiliation{Joint Institute for Nuclear Astrophysics—Center for the Evolution of the Elements (JINA-CEE), USA}

\author[0000-0003-4573-6233]{Timothy C. Beers}
\affiliation{Department of Physics, University of Notre Dame, Notre Dame, IN 46556, USA}
\affiliation{Joint Institute for Nuclear Astrophysics—Center for the Evolution of the Elements (JINA-CEE), USA}

\author[0000-0003-4479-1265]{Vinicius M. Placco}
\affiliation{Department of Physics, University of Notre Dame, Notre Dame, IN 46556, USA}
\affiliation{Joint Institute for Nuclear Astrophysics—Center for the Evolution of the Elements (JINA-CEE), USA}

\begin{abstract}
We present an examination of the metallicity distribution function of the outermost stellar halo of the Galaxy based on an analysis of both local (within 4\,kpc of the Sun, $\sim$16,500 stars) and non-local ($\sim$21,700 stars) samples. These samples were compiled using spectroscopic metallicities from the Sloan Digital Sky Survey and photometric metallicities from the SkyMapper Southern Survey. We detect a negative metallicity gradient in the outermost halo ($r >$ 35\,kpc from the Galactic center), and find that the frequency of very metal-poor ([Fe/H] $<$ $-2.0$) stars in the outer-halo region reaches up to $\sim$60\% in our most distant sample, commensurate with previous theoretical predictions. This result provides clear evidence that the outer-halo formed hierarchically. The retrograde stars in the outermost halo exhibit a roughly constant metallicity, which may be linked to the accretion of the Sequoia progenitor. In contrast, prograde stars in the outermost halo exhibit a strong metallicity-distance dependence, indicating that they likely originated from the accretion of galaxies less massive than the Sequoia progenitor galaxy.\\
\end{abstract}

\accepted{March 2, 2020}
\submitjournal{\apj}

\section{Introduction}\label{sec:intro}

Metal-poor stars have been widely recognized as ``fossils" of the earliest generation of stars in the Universe and as important tracers of the assembly history of our Galaxy. In particular, the most metal-poor stars are likely to be among the most ancient, possibly even true second-generation stars \citep[e.g.,][]{beers_2005,frebel_2015,hansen_2016,hartwig_2018}. Discerning the origins of different metal-poor populations in the Milky Way is crucial for understanding when, where, and how different Galactic components formed. Consequently, our understanding of the assembly history of the Milky Way has evolved significantly over the past half century, in large part through studies of metal-poor stars. 

\subsection{Metallicity Gradient}\label{subsec:gradient}

Many cosmological simulations suggest that galactic formation has a hierarchical assembly component and a complex merger history. \cite{amorisco_2017} used a suite of merger simulations to observe the effect of satellite mass on post-merger kinematics, and found that less-massive satellites are more likely to deposit their stars farther out in their host galaxy than more-massive satellites. It follows that the outer part of the Galactic halo may have been assembled primarily from less-massive satellites that were not able to sink deeply into the Galaxy. The least-massive satellites, which are likely the most metal-poor due to truncated star formation, may remain at the outskirts of the halo, exhibiting a trend of negative metallicity gradient with distance. The existence of such a metallicity gradient is clearly suggested in the results of \citet{starkenburg_2017}, who show that the fraction of the most metal-poor stars in galactic halos from the APOSTLE hydrodynamical simulations increases with distance. There are also several theoretical studies that assert a strong relative population of very metal-poor (VMP; [Fe/H] $< -2.0$) stars at large distances from the Galactic center, a pattern which could support the potential presence of a metallicity gradient. For instance, \cite{salvadori_2010} used high-resolution N-body simulations of a Milky Way-analogue galaxy and a semi-analytic model to analyze the  metallicity distribution function (MDF) of metal-poor halo stars, finding the relative contribution of VMP stars at distances from the Galactic center $r$ $>$ 20\,kpc to exceed 40\%. Similarly, \cite{tissera_2014} used a suite of six high-resolution Milky Way-mass systems from the Aquarius simulation project to examine the transition between the inner- and outer-halo (see Section \ref{subsec:dual_halo} below), and showed a 60\% VMP contribution to the outer-halo population, with 60\%--90\% of VMP stars coming from their simulated low-mass ($<$$10^9$\,M$_{\odot}$) satellites. 

There has also been some observational evidence presented for the existence of a negative metallicity gradient with distance. \cite{fernandez_alvar_2015} used a sample of Sloan Digital Sky Survey (SDSS; \citealt{york_2000}) stars, consisting of $\sim$1,100 stars from the Baryon Oscillations Spectroscopic Survey (BOSS; \citealt{BOSS}) and $\sim$2,800 stars from the Sloan Extension for Galactic Understanding and Exploration (SEGUE; \citealt{yanny_2009}), to demonstrate a metallicity gradient with a steep slope over Galactocentric distance $r = 20$--$40$\,kpc that flattens out at greater distances. \cite{lee_2017} demonstrated a metallicity gradient in $R$ and $|Z|$ (projected distance and height from the Galactic plane, respectively) extending up to 14\,kpc with a sample of $\sim$105,000 main-sequence turnoff (MSTO) stars from SEGUE and BOSS. \cite{yoon_2018} provide confirmation for this trend in the Southern Hemisphere, showing metallicity gradients extending up to 25\,kpc from the Galactic center with a sample of $\sim$70,000 stars from the AAOmega Evolution of Galactic Structure (AEGIS) survey. In addition to highlighting structural features in their spatial metallicity maps as evidence of the complex Galactic merger history, the latter two studies also clearly show an overall negative metallicity gradient with distance.

\subsection{Galactic Dual Halo Formation History}\label{subsec:dual_halo}

It is important to consider the effect that the stochastic merger history of the Galaxy may have on any investigations of the halo MDF. The Galactic hierarchical assembly history is complex, as evidenced by the numerous substructures discovered in the Milky Way. \cite{ELS_1962} presented a rapid, monolithic collapse scenario for the formation of the Galaxy that has been challenged by numerous studies as additional data have become available, beginning with \cite{searle_zinn_1978}. The presence of various chemodynamically distinct stellar populations \citep[e.g.,][]{yoshii_1982,gilmore_1983,carollo_2007,belokurov_2018,myeong_2018,helmi_2018,myeong_2019,matsuno_2019,carollo_2019} and evidence of various substructures \citep{ibata_1994,belokurov_2007a,belokurov_2007b} have illuminated a rich and complex Galactic formation history.

One important and surprising discovery was that the stellar halo comprises at least two distinct Galactic components: an inner, mildly prograde, more metal-rich ([Fe/H] $\sim$ $-1.6$) component and an outer, strongly retrograde, more metal-poor ([Fe/H] $\sim$ $-2.2$) component \citep[e.g.,][]{carollo_2007,carollo_2010,dejong_2010,beers_2012,an_2013,chen_2014,fernandez_alvar_2015, lee_2017, yoon_2018}.

Recent data releases of high-precision astrometric data from the Gaia mission \citep{gaia,gaia_dr2} have enabled characterization of these halo populations in great detail. Several studies have asserted that the majority of inner-halo stars were imported from a single, massive (M$_*$ $\sim$ 10$^8$--10$^9$\,M$_\odot$) progenitor that merged with the Milky Way $\sim$10 Gyr ago, known variously as the Gaia Sausage \citep{belokurov_2018,myeong_2018} or Gaia-Enceladus \citep{helmi_2018}. A somewhat less-massive (M$_*$ $\sim$ 10$^7$\,M$_\odot$) merger, dubbed the ``Sequoia Event", has also been identified as a major contributor of high-energy, retrograde, outer-halo stars \citep{myeong_2019,matsuno_2019}.\\

Motivated by these recent advances in our understanding of the complex Galactic assembly history, we seek to further investigate the existence of a metallicity gradient in the ``outermost halo" and to consider the complex formation history of this component. While the result by \cite{fernandez_alvar_2015} shows a metallicity gradient over a larger distance range than \citet{lee_2017} and \citet{yoon_2018}, the number of stars considered is rather small ($\sim$4,000) compared to the latter studies ($\lesssim$100,000). These results could also be accounted for by the overlapping inner- and outer-halo populations, and the gradual shift in the relative dominance of these components with increasing distance from the Galactic center. More importantly, one shortcoming of the existing observational studies is the possible presence of a metallicity-distance selection bias\textbf{---}more metal-poor giants (the dominant class in distant samples) are brighter than their more metal-rich counterparts, so one is more likely to select more metal-poor stars when observing the outer-halo region, which could artificially induce a gradient in subsequent analyses. 

In this work, we suggest the presence of a negative metallicity gradient in the outer-halo's MDF over $r$, using non-local samples (``{\it in-situ}"). More importantly, to mitigate the metallicity-distance bias problem, we also perform our analyses with local samples (within 4\,kpc of the Sun). These local samples allow us to observe a definitive metallicity gradient over apocentric distance, $r_{apo}$ (``{\it ex-situ}"). We introduce our {\it in-situ} and {\it ex-situ} samples in Section \ref{sec:data} and describe our kinematic analyses of these samples in Section \ref{sec:kin}. In Section \ref{sec:results_disc}, we discuss our two important findings: 1) a metallicity gradient does indeed exist at large distances ($>$35\,kpc) in the halo, particularly in the prograde direction and 2) retrograde stars appear to possess a flat metallicity-distance relation, indicating that the progenitor of the retrograde outer-halo is likely associated with the Sequoia merger event.  Finally, we summarize our results and discuss potential future investigations of the outermost halo's MDF in Section \ref{sec:summ_conc}.

\section{Data}\label{sec:data}

\begin{figure*}
\plottwo{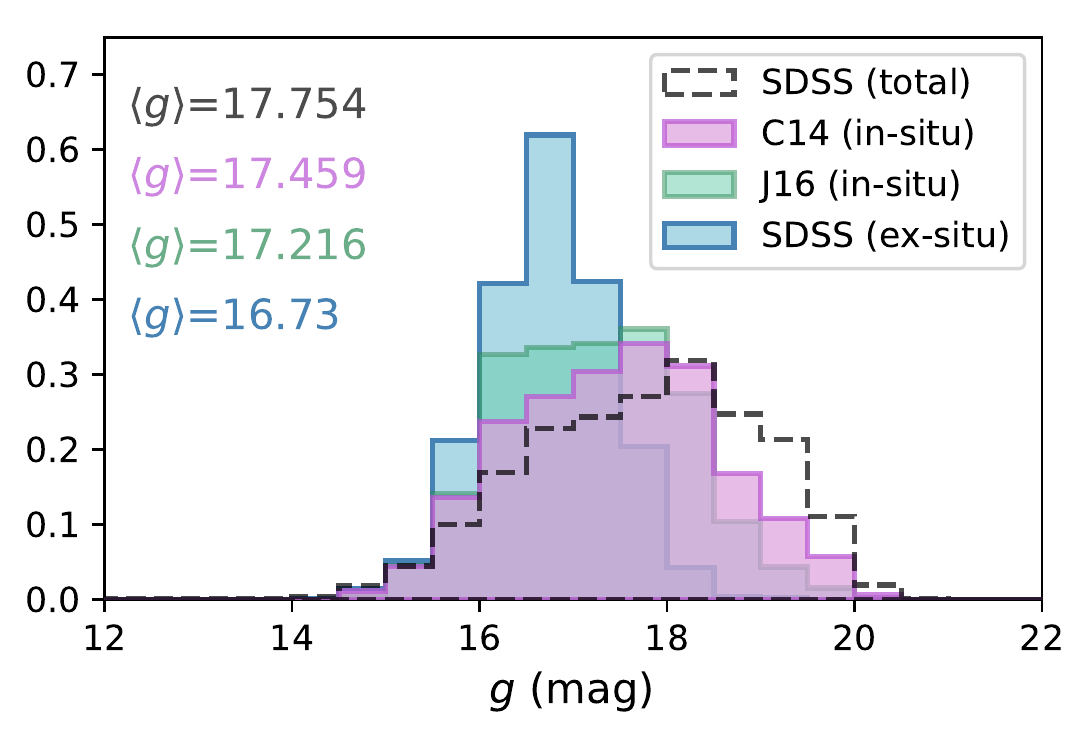}{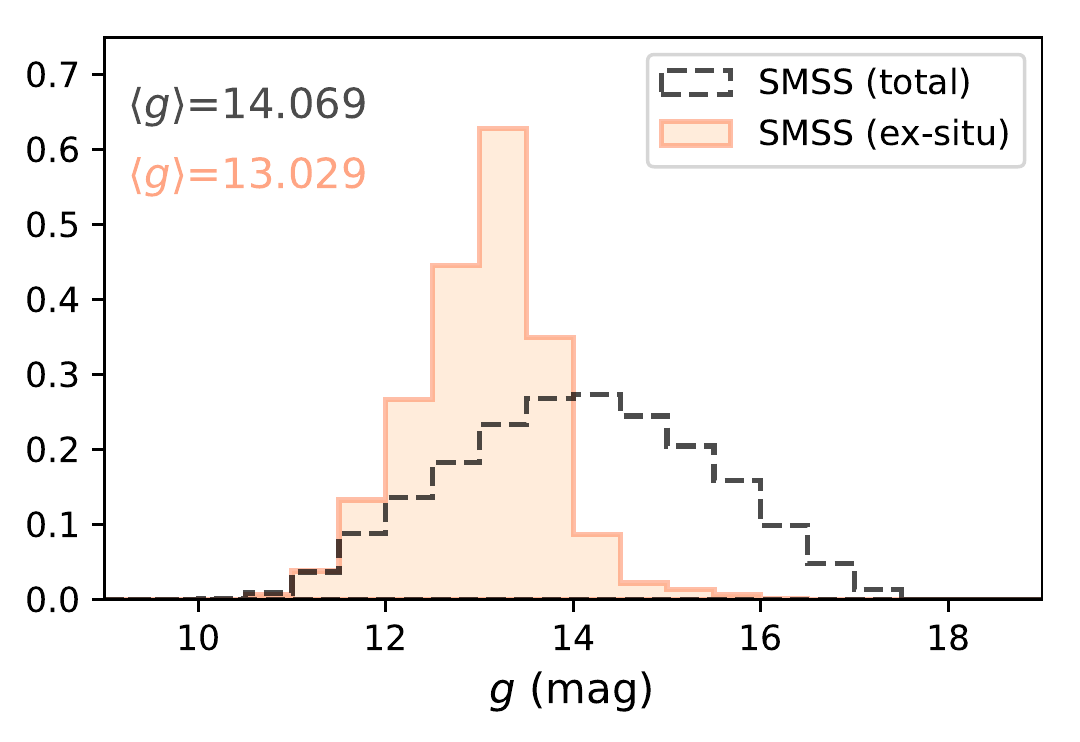}
\caption{\textbf{Normalized magnitude distributions in the $g$-band for all samples. The three SDSS datasets (the {\it in-situ} C14 and J16 samples and the {\it ex-situ} SDSS sample) are grouped together in the left panel. The {\it ex-situ} SMSS sample is shown in the right panel. In each panel, the total sample (without cuts on location or kinematics) is shown with a dashed line for reference. Mean $g$-band magnitudes ($\langle g \rangle$) are noted in the upper left of each panel.} \label{fig:g_distribs}}
\end{figure*}

\begin{figure}
\plotone{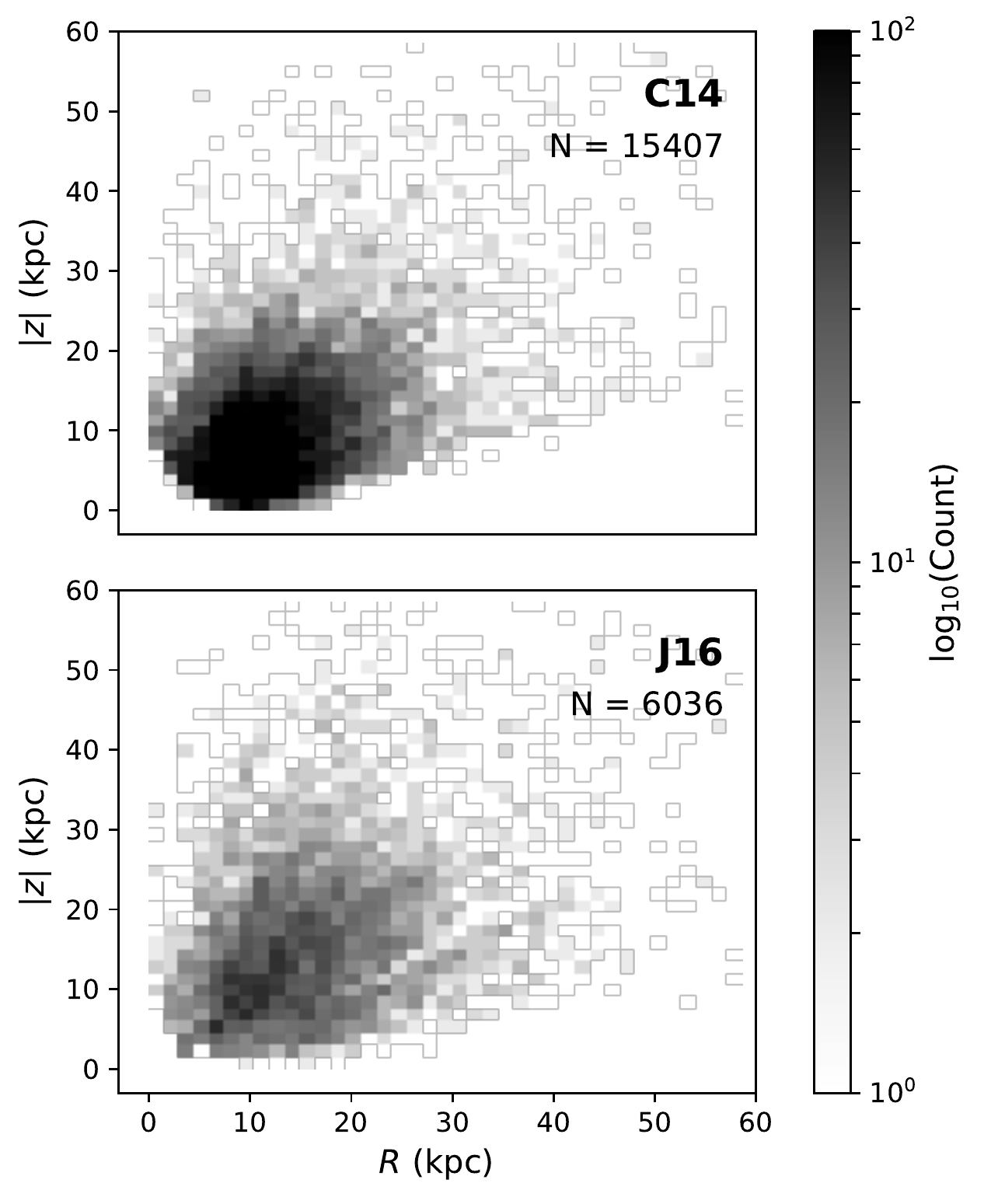}
\caption{\textbf{Spatial distributions of stars with valid kinematics for our two in-situ samples (C14 and J16), represented in a Galactocentric reference frame. Projected Galactocentric radius onto the Galactic plane, $R$, is shown on the x-axis. Vertical distance from the Galactic plane, $|z|$, is shown on the y-axis. Counts per bin are represented via a logarithmic scale. Note that some stars ($<$1.5\% per sample) lie outside the spatial axis boundaries used in these plots.} \label{fig:R_z}}
\end{figure}

\subsection{Non-Local ``{\it In-situ}"  Samples}\label{subsec:in_situ}

Our {\it in-situ} samples consist of two sets of SDSS \textbf{($R \sim$ 2,000)} giants compiled by \citet{chen_2014} and \citet{janesh_2016}. The \citet[][hereafter, C14]{chen_2014} sample comprises 15,723 red giant branch (RGB) stars from SDSS DR9 \citep{sdss_dr9}, compiled to study the thick disk, inner-halo, and outer-halo of the Galaxy. The \citet[][hereafter, J16]{janesh_2016} sample is made up of 6,036 K giants from SDSS DR9, selected to study substructure in the stellar halo. Both samples cover distance ranges in excess of 100\,kpc, making them good candidate data-sets for studying the outermost halo. \textbf{The majority of stars in C14 and J16 do not have reliable Gaia parallaxes (with $<$20\% uncertainty) available. We use the original distances derived in \citet{chen_2014} for C14, and calculate photometric distances for J16 following the method described in \citet{beers_2000}. We also retain the original SDSS radial velocities (mean uncertainty $\sim$ 2\,km\,s$^{-1}$) and proper motions for both samples.
Spatial distributions for these samples are included for reference in Figure \ref{fig:R_z}. The C14 sample is highly concentrated within $\sim$20\,kpc of the Galactic center while the J16 sample is distributed more uniformly. Though the majority of both samples lies within $r \sim 50$\,kpc, each spans a range of $>$100\,kpc.}
 
\subsection{Local ``{\it Ex-situ}" Samples}\label{subsec:ex_situ}

The primary {\it ex-situ} sample used for our analyses is compiled from SDSS DR15 \citep{sdss_dr15}. The initial \textbf{query to the SDSS catalog server\footnote{https://skyserver.sdss.org/casjobs/}} resulted in 357,816 stars with signal-to-noise ratios (SNR) $>$10 in the effective temperature range 4,500\,K $<$ $T_{\rm eff}$ $<$ 7,000\,K, where the SEGUE Stellar Parameter Pipeline \citep[SSPP;][]{lee_2008} is most reliable. Duplicate stars were removed by choosing the measurement with the highest SNR. Stars with spectra taken on plug-plates which were part of SEGUE cluster- or structure-targeting programs were removed prior to analyses\footnote{See ``SEGUE Target Selection" on the SDSS DR15 website for details (www.sdss.org/dr15/algorithms/segue\_target\_selection).}.

Proper motions and radial velocities were taken from Gaia DR2  (\citealt{gaia_dr2}) where available. When unavailable, we used the original kinematic parameters from the SDSS archive. \textbf{The resulting sample contains mainly ($>$99\%) {\it Gaia} proper motions and SDSS radial velocities.} We adopt distances from the Bailer-Jones treatment of the Gaia parallaxes, and restrict the sample to stars with $<$20\% distance uncertainty \citep{bailer_jones_2018}.
Stars with uncertainties on their radial velocities exceeding 20\,km\,s$^{-1}$ were removed from the sample. \textbf{The resulting sample has a mean radial velocity uncertainty of $\sim$ 1.5\,km\,s$^{-1}$.}
We limit our sample to a local volume within 4\,kpc of the Sun, leaving us with 118,037 stars.

Our complementary {\it ex-situ} sample was compiled by \cite{huang_2019} from DR1 of the SkyMapper Southern Survey (SMSS; \citealt{wolf_2018}). The authors provide metallicity estimates for 972,994 RGB stars and compile kinematic parameters where available. Proper motions and \textbf{(Bailer-Jones)} distances  from Gaia DR2 are available for the majority \textbf{($\sim$70\%)} of the sample. Of the 972,994 stars in this sample, 423,995 have available radial-velocity estimates, \textbf{compiled by \citet{huang_2019}} from a  variety of catalogues, primarily Gaia DR2 and the Galactic Archaeology with HERMES survey (GALAH\textbf{, $R \sim$ 28,000}; \citealt{galah_dr2}). \textbf{The resulting sample has a mean radial velocity uncertainty of $\sim$ 1 km\,s$^{-1}$.} After limiting the sample to a local volume, we are left with a sample of 395,144 stars for which viable kinematics can be obtained.

\textbf{A comparison of the $g$-band magnitude distributions for all samples used in our analyses is shown for reference in Figure \ref{fig:g_distribs}. We divide our samples into two plots based on survey source, as the SDSS $g$-band and the SMSS $g$-band differ from each other\footnote{http://skymapper.anu.edu.au/filter-transformations/} and cover a different magnitude range.}

\section{Kinematical Analysis}\label{sec:kin}

Kinematic parameters, such as $U$, $V$, and $W$, are derived using the \texttt{galpy} Python package \citep{galpy}. In this work we adopt $R_\odot$ = 8\,kpc as the Sun's distance from the Galactic center projected onto the Galactic plane. We use $v_{LSR}$ = 220\,km\,s$^{-1}$ for the velocity of the local standard of rest (LSR) \citep{kerr_1986} and ($U$, $V$, $W$)$_\odot$ = ($-$9, 12, 7)\,km\,s$^{-1}$ \citep{mihalas_1981} for the motion of the Sun with respect to the LSR.

Orbital parameters such as energy, maximum height from the Galactic plane ($Z_{max}$), and $r_{apo}$ are derived with a version of the potential code used in \cite{chiba_2000}, adapted to run in Python. This code adopts the analytic St\"ackel potential developed by \cite{sommer-larsen_1990}, consisting of a flattened, oblate disk and a nearly spherical massive halo. We note that we use the St\"ackel potential and not \texttt{galpy's} Milky Way-like potential, \texttt{MWPotential2014}, following the recent comparison of \cite{kim_2019} between the St\"ackel and \texttt{galpy} potential. \cite{kim_2019} suggest that \texttt{MWPotential2014} may not be the ideal choice for studies focusing on the outer-halo because many highly energetic, predominantly retrograde (outer-halo) stars are found to be unbound when using the shallower \texttt{galpy} potential.

We estimate uncertainties on orbital values produced by the St\"ackel potential via a Markov Chain Monte Carlo sampling method. Assuming that the uncertainties on the input parameters are normally distributed about their observed values, we generate 1,000 randomly sampled orbits per star, and adopt the standard deviation of the resulting distribution as our uncertainty. Since our {\it in-situ} non-local samples do not utilize high-precision Gaia data, they have larger uncertainties on average than our {\it ex-situ} samples. We choose not to make any cuts on uncertainty for C14 and J16 because such cuts would exclude an overly large number of stars from our analyses. Our two {\it ex-situ} samples utilize Gaia data, so we are able to trim high-uncertainty stars from our data to minimize potentially spurious features. 

\textbf{The manner in which we prune high-uncertainty stars from our samples depends on the subsequent analyses we intend to perform on them.}
In Section \ref{subsec:ex_situ_results} we bin stars over $r_{apo}$ in steps of size 5\,kpc each\textbf{, so w}e choose to restrict our sample to stars with uncertainty on $r_{apo}$ $<$ $\pm$10\,kpc\textbf{. T}his allows us to create a low-uncertainty sample without losing too many distant halo stars due to an overly strict cutoff. To minimize contamination from disk-system stars in our local samples, we exclude stars with $r_{apo}$ $<$ 10\,kpc and $Z_{max}$ $<$ 3\,kpc, leaving us with 10,078 SDSS halo stars and 6,576 SMSS halo stars. 

\textbf{Select values for our compiled \textit{ex-situ} samples are available in Tables \ref{tab:SDSS} and \ref{tab:SMSS}. The final \textit{ex-situ} samples' derived orbits span distances up to $Z_{max}\sim 60$\,kpc and $r_{apo}\sim 80$\,kpc.}

\begin{splitdeluxetable*}{cccrrrrrBrrrrr}
\tablecaption{Select compiled parameters and derived quantities for our {\it ex-situ} SDSS sample. \label{tab:SDSS}}
\tablehead{
\colhead{Plate} & \colhead{MJD} & \colhead{Fiber} & \colhead{[Fe/H]} &  \colhead{ra} & \colhead{dec} & \colhead{pmra} & \colhead{pmdec} & \colhead{RV} & \colhead{dist} & \colhead{$v_\phi$} & \colhead{$r_{apo}$} & \colhead{$Z_{max}$}\\
\colhead{} & \colhead{} & \colhead{} & \colhead{} & \colhead{(deg)} & \colhead{(deg)} & \colhead{(mas~yr$^{-1}$)} & \colhead{(mas~yr$^{-1}$)} & \colhead{(km~s$^{-1}$)} & \colhead{(kpc)} & \colhead{(km~s$^{-1}$)} & \colhead{(kpc)} & \colhead{(kpc)}
}
\startdata
51930 & 489 & 474 & $-$1.376 & 158.523 & 65.498 & 3.624 $\pm$ 0.090 & $-$8.199 $\pm$ 0.098 & $-$249.304 $\pm$ 3.268 & 3.521 $\pm$ 0.648 & 49.047 & 16.040 & 4.506 \\
51930 & 489 & 245 & $-$1.652 & 157.353 & 64.309 & $-$18.414 $\pm$ 0.085 & $-$13.070 $\pm$ 0.102 & $-$80.675 $\pm$ 3.250 & 1.860 $\pm$ 0.214 & 54.139 & 10.099 & 4.524 \\
51882 & 442 & 504 & $-$1.549 & 126.276 & 51.224 & $-$11.012 $\pm$ 0.077 & $-$25.211 $\pm$ 0.065 & 114.838 $\pm$ 2.487 & 2.361 $\pm$ 0.309 & $-$15.414 & 15.967 & 3.897 \\
51882 & 442 & 161 & $-$1.695 & 125.581 & 50.654 & 7.362 $\pm$ 0.076 & $-$36.842 $\pm$ 0.057 & $-$211.139 $\pm$ 2.195 & 1.291 $\pm$ 0.083 & $-$25.832 & 12.116 & 3.020 \\
53240 & 1894 & 395 & $-$1.704 & 354.664 & 15.054 & 34.733 $\pm$ 0.094 & $-$5.482 $\pm$ 0.057 & $-$68.647 $\pm$ 2.637 & 1.981 $\pm$ 0.202 & $-$17.872 & 16.628 & 3.141 \\
\enddata
\tablecomments{Table \ref{tab:SDSS} is published in its entirety in the machine-readable format. A portion is shown here for guidance regarding its form and content.}
\end{splitdeluxetable*}

\begin{splitdeluxetable*}{crrrrrBrrrrr}
\tablecaption{Select compiled parameters and derived quantities for our {\it ex-situ} SMSS sample. \label{tab:SMSS}}
\tablehead{
\colhead{SMSSID} & \colhead{[Fe/H]} &  \colhead{ra} & \colhead{dec} & \colhead{pmra} & \colhead{pmdec} & \colhead{RV} & \colhead{dist} & \colhead{$v_\phi$} & \colhead{$r_{apo}$} & \colhead{$Z_{max}$}\\
 \colhead{} & \colhead{} & \colhead{(deg)} & \colhead{(deg)} & \colhead{(mas~yr$^{-1}$)} & \colhead{(mas~yr$^{-1}$)} & \colhead{(km~s$^{-1}$)} & \colhead{(kpc)} & \colhead{(km~s$^{-1}$)} & \colhead{(kpc)} & \colhead{(kpc)}
}
\startdata
222786568 & $-$1.984 & 267.648 & $-$58.718 & 1.454 $\pm$ 0.036 & 0.803 $\pm$ 0.040 & $-$122.650 $\pm$ 1.980 & 12.264 $\pm$ 1.017 & $-$98.215 & 29.857 & 19.887 \\
222773278 & $-$1.629 & 267.168 & $-$58.101 & 3.696 $\pm$ 0.059 & $-$7.926 $\pm$ 0.062 & $-$131.800 $\pm$ 2.320 & 5.389 $\pm$ 0.448 & 266.741 & 11.134 & 4.179 \\
134079039 & $-$1.645 & 286.969 & $-$26.781 & $-$0.112 $\pm$ 0.044 & $-$7.878 $\pm$ 0.036 & $-$313.540 $\pm$ 2.880 & 10.782 $\pm$ 0.795 & $-$16.258 & 16.315 & 12.479 \\
134123378 & $-$2.680 & 287.205 & $-$26.187 & $-$0.153 $\pm$ 0.078 & $-$18.165 $\pm$ 0.070 & $-$250.420 $\pm$ 2.570 & 5.906 $\pm$ 0.136 & $-$337.081 & 15.519 & 6.025 \\
134322209 & $-$1.157 & 287.935 & $-$25.698 & $-$2.178 $\pm$ 0.063 & 2.817 $\pm$ 0.060 & $-$70.780 $\pm$ 1.490 & 6.603 $\pm$ 1.581 & 198.729 & 10.090 & 5.753 \\
\enddata
\tablecomments{Table \ref{tab:SMSS} is published in its entirety in the machine-readable format. A portion is shown here for guidance regarding its form and content.}
\end{splitdeluxetable*}

\begin{figure*}
\epsscale{0.575}
\plotone{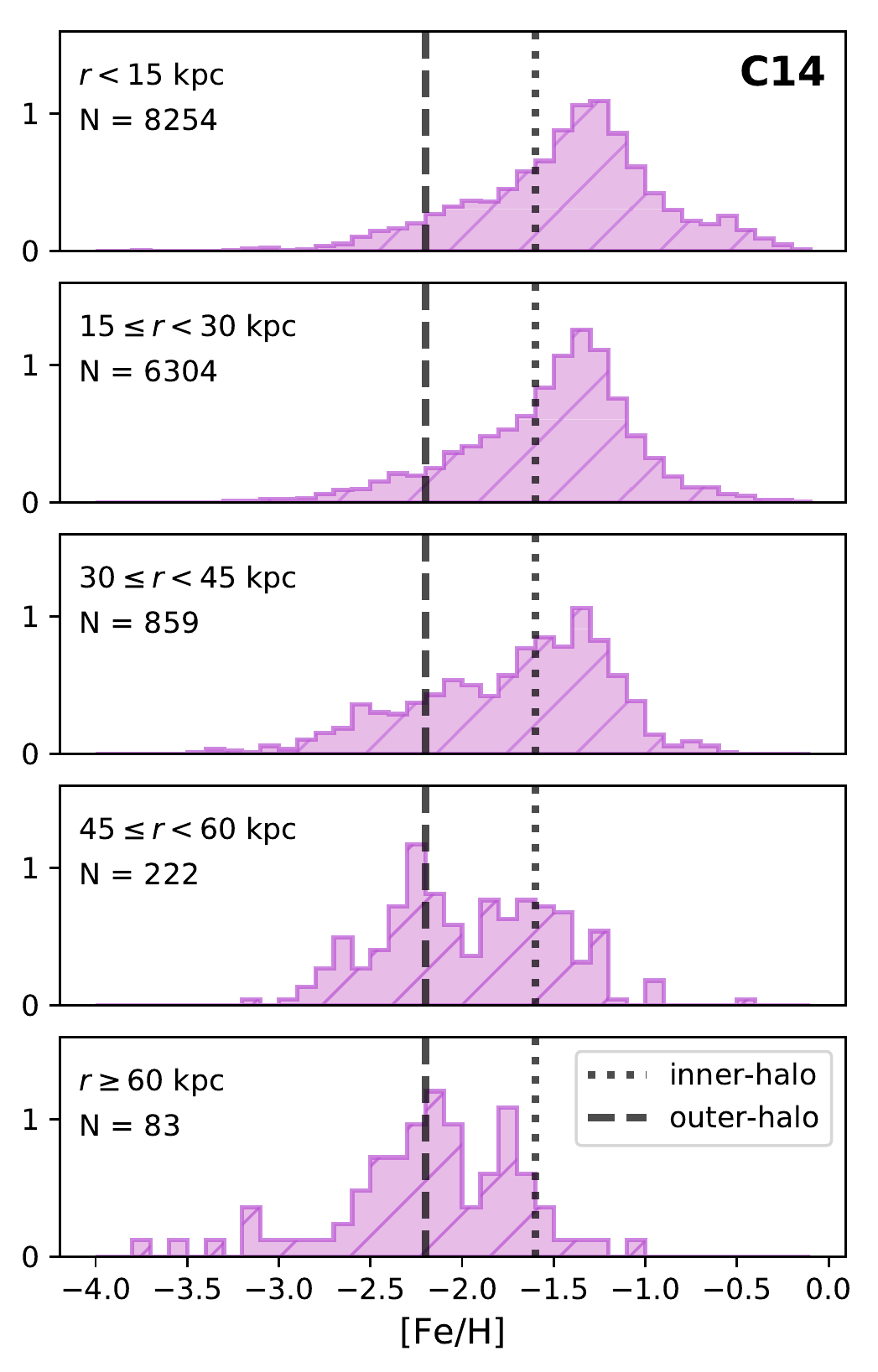}
\plotone{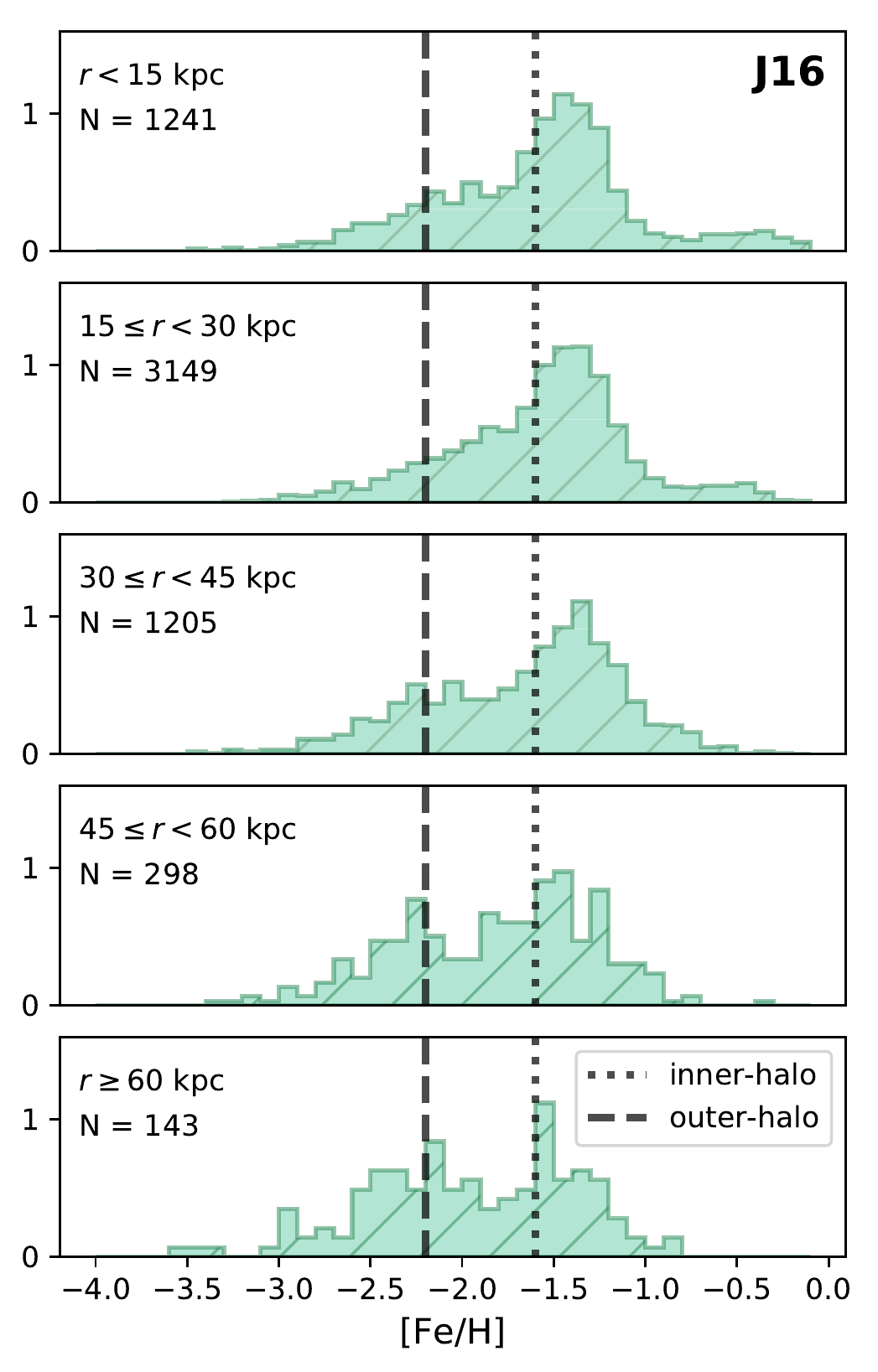}
\caption{Metallicity distributions for the C14 (left column) and J16 (right column) samples over increasing slices of $r$. The dotted and dashed lines mark the mean metallicities of the inner-halo ([Fe/H] = $-1.6$) and outer-halo ([Fe/H] = $-2.2$), respectively, estimated by \cite{carollo_2010}.  Note the apparent increase in the tail strength of the MDFs at low metallicity in the lower (more distant) panels, beginning around $r$ $>$ 30\,kpc.  See text for discussion. \label{fig:C14_J16}}
\end{figure*}

\section{Results and Discussion}\label{sec:results_disc}

\subsection{Global Metallicity Gradient}\label{subsubsec:global}

\subsubsection{{\it In-situ} Results}\label{subsec:in_situ_results}

In Figure \ref{fig:C14_J16} we construct MDFs for C14 (left panels) and J16 (right panels) over increasing slices of $r$The MDF of the C14 sample shifts toward the metal-poor regime and its metal-poor tail noticeably increases in relative proportion as we move farther from the Galactic center, particularly beyond 45\,kpc. Though this effect is not as noticeable in the J16 sample as the C14 sample, likely due to the J16 selection function\footnote{In an effort to remove foreground dwarfs from their sample, these authors trimmed stars with spectra having significantly strong MgH features. Unfortunately, this also {\bf resulted} in the removal of a significant number of carbon-enhanced metal-poor (CEMP) stars, which are among the most likely to be VMP stars.}\textbf{,} both clearly show that the dominant stellar component changes from the metal-richer populations of the metal-weak thick disk (MWTD; [Fe/H] $\sim$ $-0.8$ to $-1.8$) and inner-halo ([Fe/H] $\sim$ $-1.6$)  to the metal-poor outer-halo ([Fe/H] $\sim$ $-2.2$) over increasing distance. We note that there exists, interestingly, a relatively strong inner-halo like population even beyond 30\,kpc in both samples. This may be associated with a major merger event, and is discussed \textbf{in} more detail in Section~\ref{subsec:ex_situ_results}.

Although our {\it in-situ} samples are likely to be affected by metallicity selection biases, here we are interested in the nature of their {\it lowest-metallicity} tails, with [Fe/H] $<$ $-2.0$.  As clearly shown by the empirical comparison of giant-branch luminosity with metallicity for Galactic globular clusters in Figure 5 of \citet{huang_2019}, this dependency is minimal at the lowest abundances. Additionally, the increasingly apparent bimodality at larger distances in Figure \ref{fig:C14_J16} (also seen in \citealt{carollo_2007,carollo_2010}) cannot be explained by this bias alone.

\subsubsection{{\it Ex-situ} Results}\label{subsec:ex_situ_results}
We conduct a similar analysis using the local {\it ex-situ} SDSS and SMSS samples to mitigate the metallicity-distance bias, based on the suggestive evidence of a possible metallicity gradient in the outermost halo from the non-local {\it in-situ} samples analyzed above. 

\begin{figure*}
\plottwo{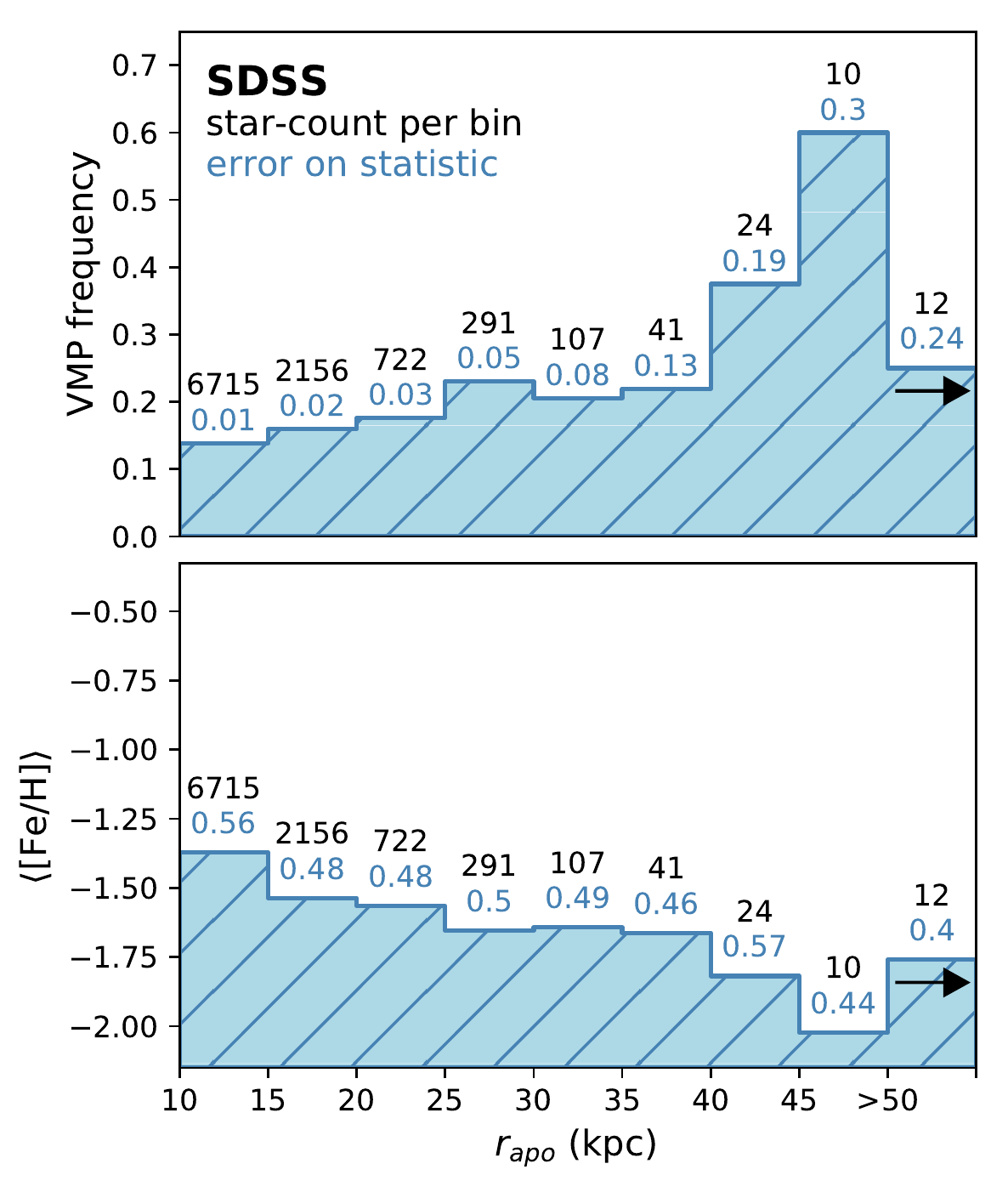}{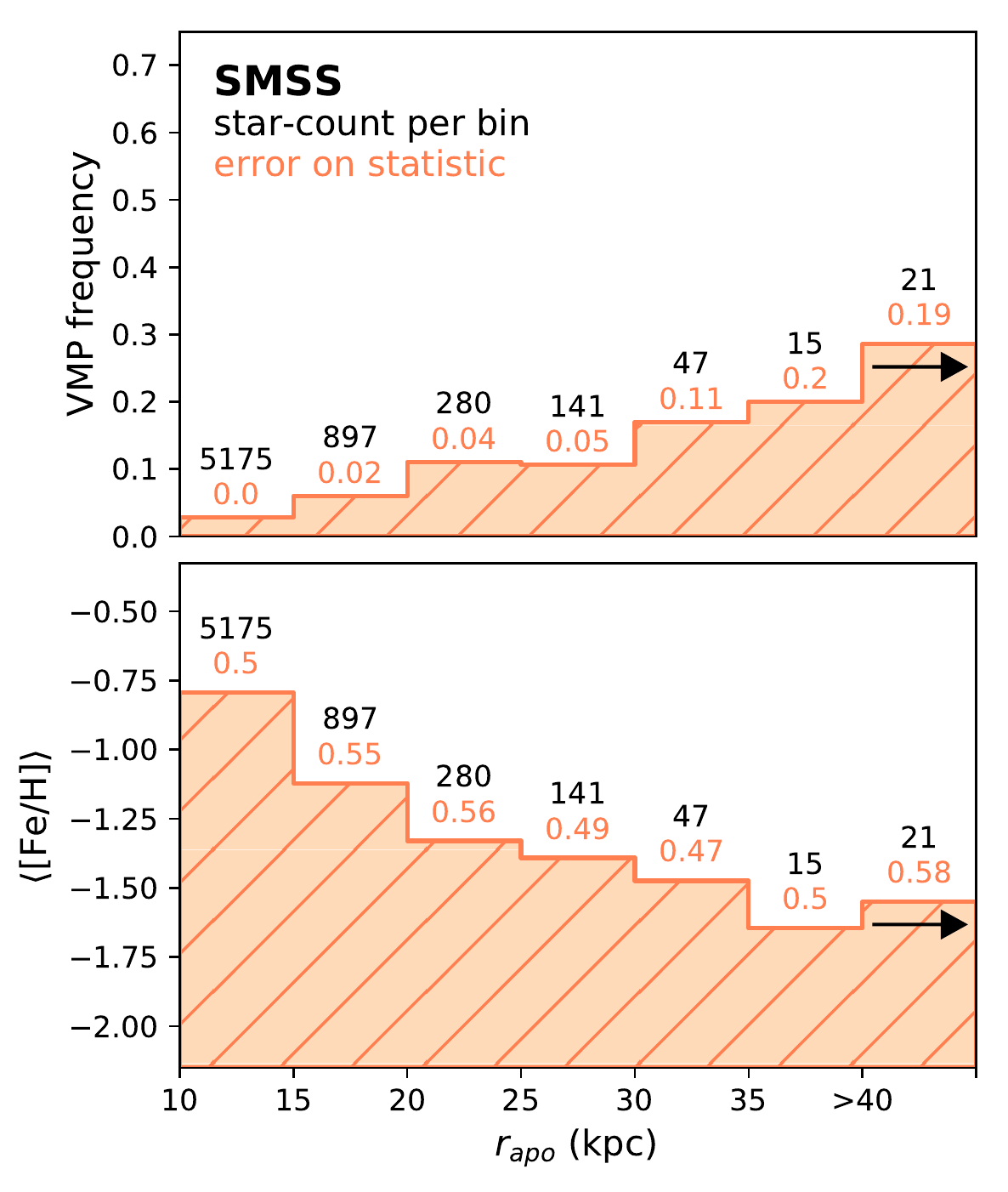}
\caption{VMP frequencies (top) and average metallicities ($\langle$[Fe/H]$\rangle$, bottom) for the SDSS (left) and SMSS (right) samples. In the upper panels, star counts and frequency error estimates are indicated for each bin. The frequency error is estimated with a one-sigma Wilson proportion confidence interval \citep{wilson_1927}.  In the lower panels, star counts and the dispersion of [Fe/H] for each bin are provided. \label{fig:all_binned}}
\end{figure*}

We examine how the frequency of VMP stars and the average metallicity of our samples vary as a function of $r_{apo}$, as shown in the upper and lower panels of Figure \ref{fig:all_binned}, respectively. Stars are binned in steps of size 5\,kpc until fewer than 10 stars are available per bin, after which all subsequent bins are combined. We note that, because of this binning choice, the final bin in each panel may be influenced by low-number statistics and suffer from high uncertainty.

The VMP frequency of the SDSS sample shown in the top left panel of Figure \ref{fig:all_binned} climbs very slowly in the range $r_{apo}$ = 10--40\,kpc, after which it experiences a sharp increase, rising from $\sim$20\% to $\sim$60\% over the next 10\,kpc. The average metallicity slowly decreases from [Fe/H] $\sim$ $-$1.4 at $r_{apo}$ = 10\,kpc, plateauing around [Fe/H] $\sim$ $-$1.6 at $r_{apo}$ = 25--40\,kpc, then dropping rapidly to [Fe/H] $\sim$ $-2.0$. We note that the statistics in the largest distance bins of the left panels appear contrary to the overall trends, but have high uncertainty compared to the majority of the preceding bins.

The VMP frequency for the SMSS photometric sample experiences a steady climb over $r_{apo}$, maxing out at $\sim$30\%, as seen in the top-right panel of Figure \ref{fig:all_binned}. The average metallicity decreases from [Fe/H] $\sim$ $-0.8$ at 10\,kpc to [Fe/H] $\sim$ $-1.6$ at 40\,kpc. Although the changes in VMP frequency and mean metallicity are not as dramatic for the SMSS sample as for the SDSS sample, both samples display a clear metallicity gradient over $r_{apo}$. The differences in the samples are likely due to the fact that the SMSS sample does not reach as far into the halo as the SDSS sample.

The steep change in mean metallicity from [Fe/H] $\sim$ $-1.6$ to \textbf{approximately} $-2.0$ at $r_{apo} \gtrsim 40$\,kpc in the SDSS panels in Figure \ref{fig:all_binned} could indicate a shift between the relative dominance of the inner-halo ([Fe/H] $\sim$ $-1.6$) and outer-halo ([Fe/H] $\sim$ $-2.2$), as discussed in Section~\ref{subsec:dual_halo}. Another possibility is that this region of the metallicity distribution represents the shift between stars donated to the outer-halo by the Sequoia merger event, estimated at [Fe/H] $\sim$ $-1.6$ \citep{myeong_2019}, and stars donated by smaller mergers of more metal-poor satellites. Accordingly, we  investigate the detailed Galactic halo assembly history using these local samples in the next subsection.

\subsection{Detailed Accretion History }

\subsubsection{Metallicity Distribution}\label{subsubsec:MDF}
\begin{figure*}[thb!]
\plottwo{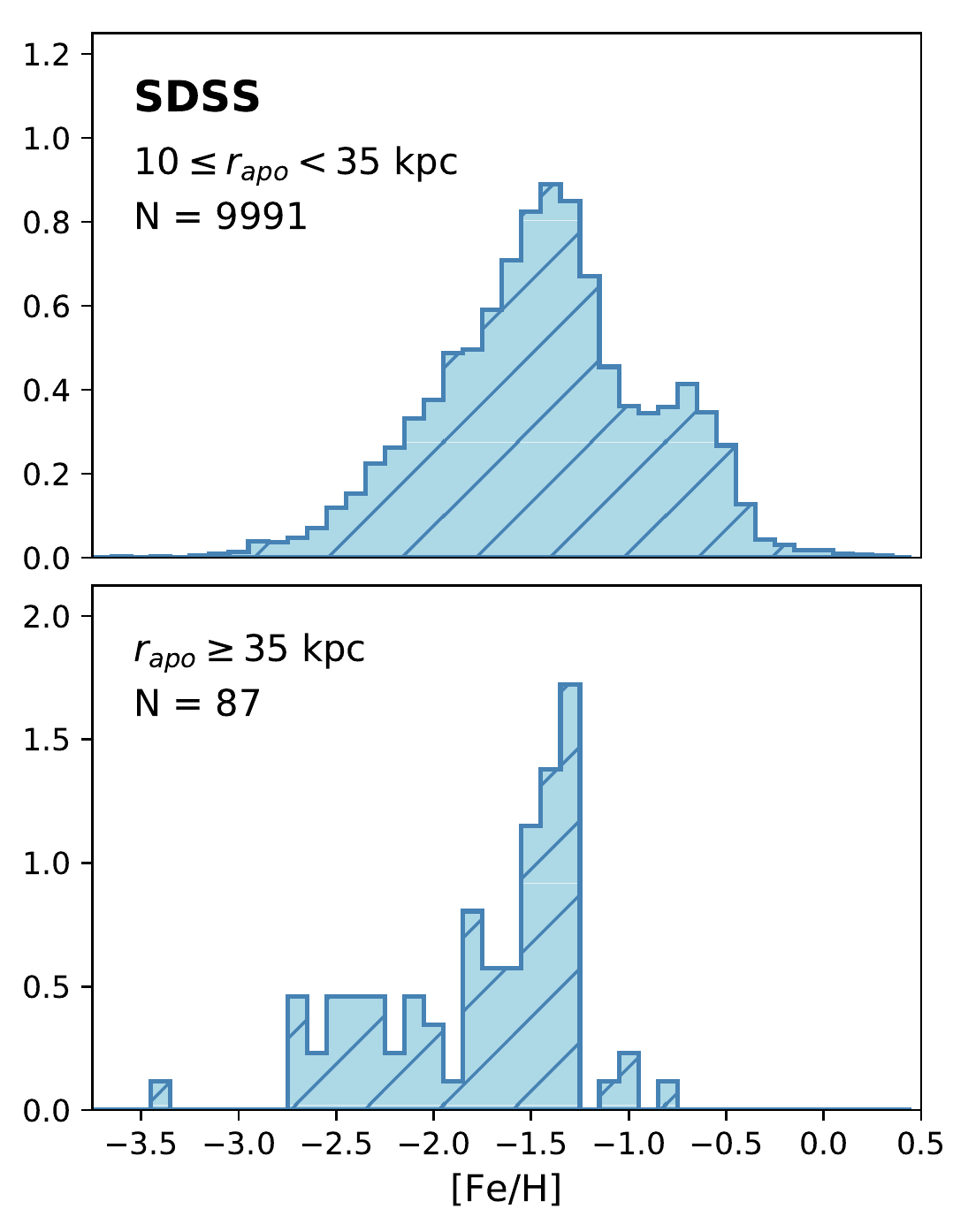}{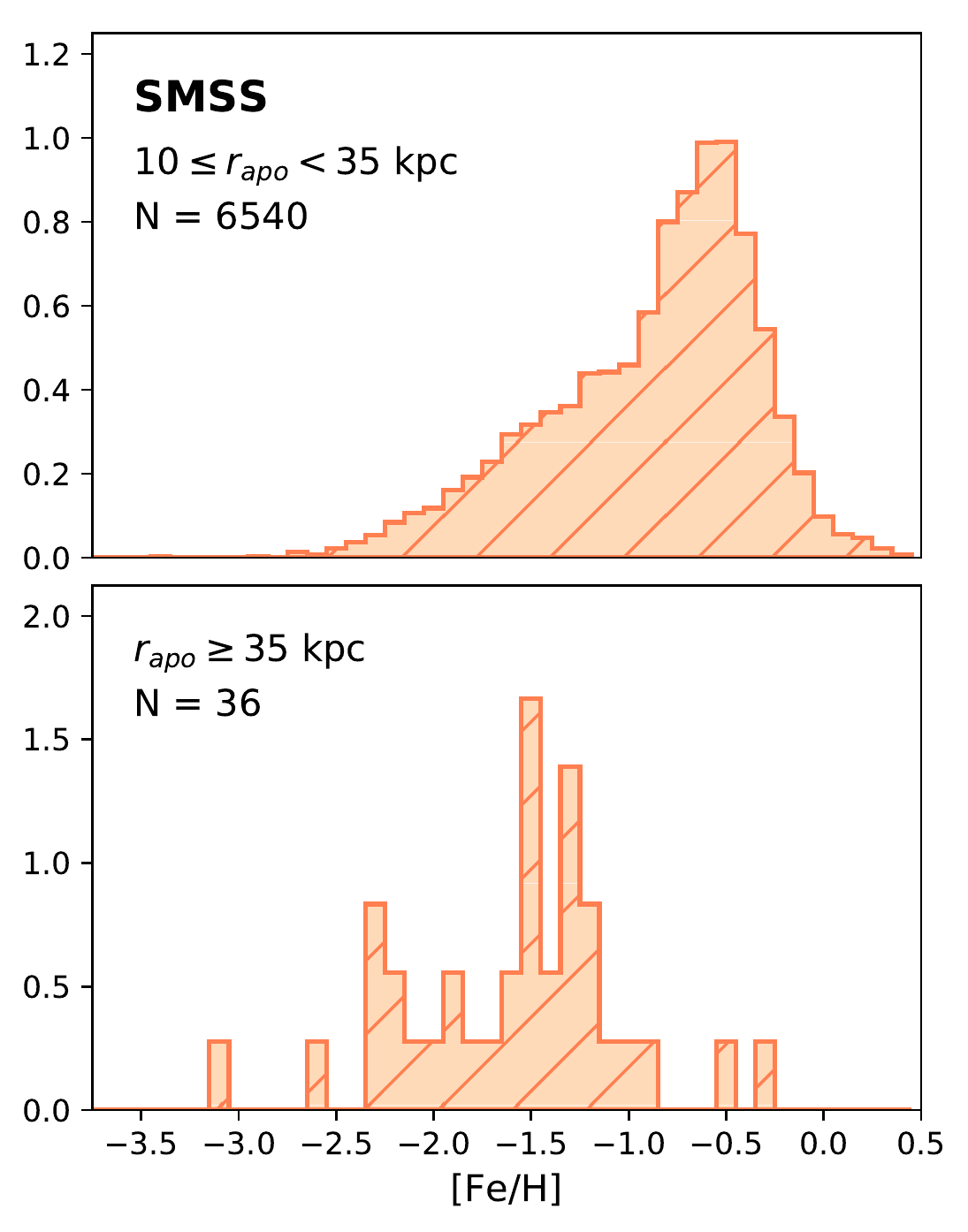}
\caption{\textbf{Normalized} MDFs for the SDSS sample (left) and the SMSS sample (right), divided at r$_{\rm apo}$ = 35\,kpc. \label{fig:MDFs}}
\end{figure*}

\begin{figure}
\plotone{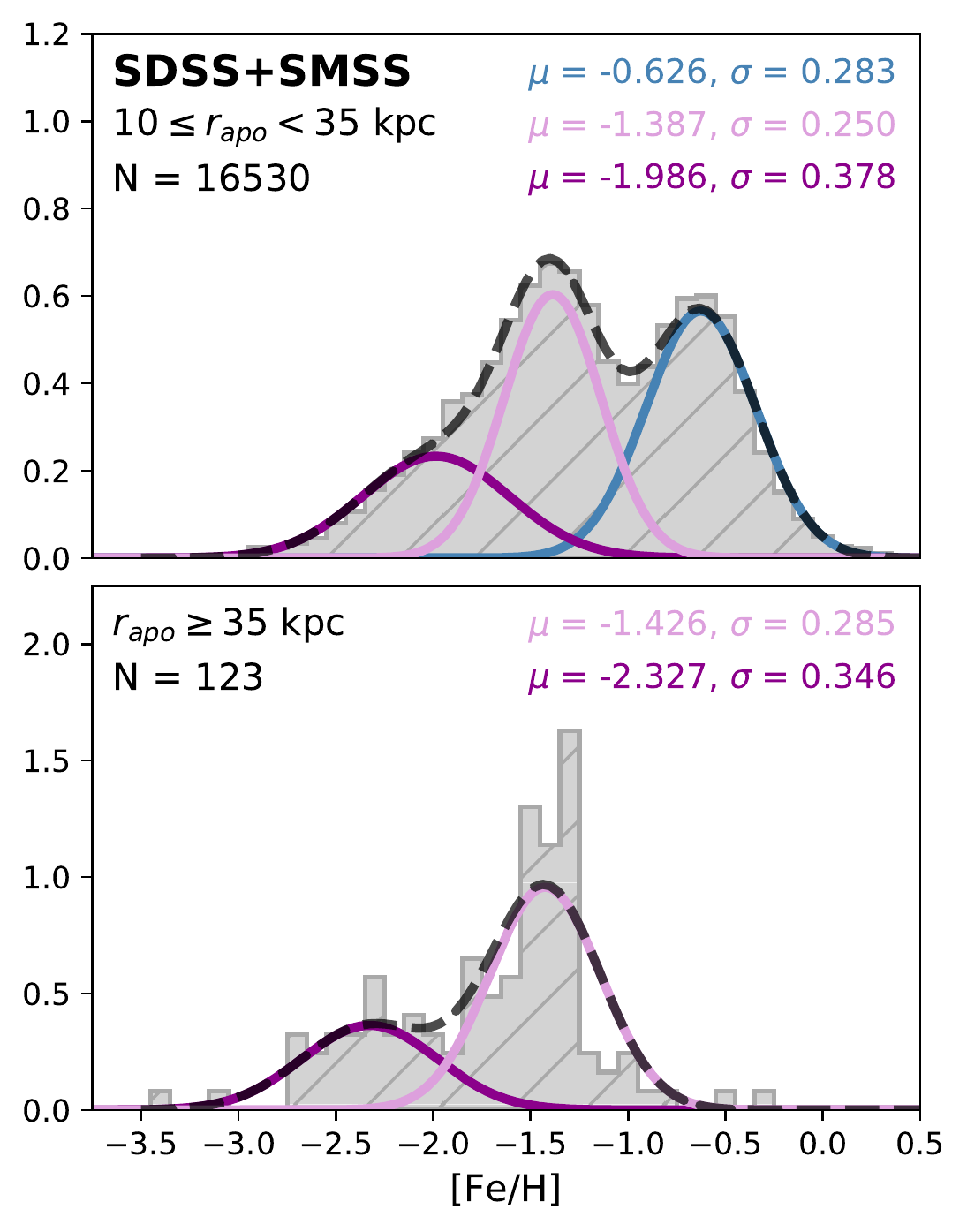}
\caption{\textbf{Normalized} MDFs for the combined (SDSS plus SMSS) sample, divided on r$_{\rm apo}$ = 35\,kpc. The color-coded curves represent Gaussian mixture-model fits of the high-r$_{\rm apo}$ (bottom) and low-r$_{\rm apo}$ (top) distributions. The dashed curves show the sum of these components. \label{fig:GMM}}
\end{figure}

Since the behavior of the SDSS local halo stars changes at approximately 35--40\,kpc (e.g., Figure \ref{fig:all_binned}, left panels), we further investigate the reason for this behavior by constructing MDFs for ``near" (10 $\leq$ $r_{apo}$ $<$ 35\,kpc) and ``far" ($r_{apo}$ $\geq$ 35\,kpc) halo samples, shown in the left panels of Figure \ref{fig:MDFs}. A similar diagram is shown for the SMSS halo stars in the right panels of Figure \ref{fig:MDFs}. The SMSS halo stars do not present the same sharp changes at 35--40\,kpc (see Figure \ref{fig:all_binned}, right panels), in part because they do not probe the same distance range as the SDSS sample. 

However, since the SMSS sample displayed similar general characteristics to the SDSS sample in our analyses of Figure \ref{fig:all_binned}, we chose to create additional near/far MDFs from the combination of both the SDSS and SMSS samples, in order to bolster the number of stars available in the $r_{apo}$ $\geq$ 35\,kpc range (see Figure \ref{fig:GMM}). Gaussian distributions are fit to the combined data using the \texttt{scikit-learn} Gaussian Mixture Model (GMM) package in Python to identify components with potentially distinct origins. The near-halo combined MDF primarily consists of a distinctive component at [Fe/H] $\sim$ $-1.4$, with a smaller, more metal-rich peak \textbf{around} $-0.6$, possibly belonging to a portion of the MWTD that was not completely removed by the cuts made in Section \ref{sec:kin}. The far-halo combined MDF also has a dominant [Fe/H] $\sim$ $-1.4$ peak, as well as a more metal-poor component at [Fe/H] $\sim$ $-2.3$. The near halo also possesses a small VMP population, fitted with a peak at [Fe/H] $\sim$ $-2.0$, but this population rises in relative significance in the far halo.

Figure \ref{fig:GMM} shows that there may be at least two separate populations at $r_{apo}$ $\geq$ 35\,kpc. The more metal-rich peak ([Fe/H] $\sim$ $-1.4$) seen in the far-halo (lower) panel of Figure \ref{fig:GMM} could be a selection of inner-halo population stars still present at $r_{apo}$ $\geq$ 35\,kpc. The exact location of the transition zone between the inner- and outer-halo regions is uncertain. \cite{carollo_2007} place it at $\sim$ 15-20\,kpc while \cite{kim_2019} give an estimate of $\sim$30\,kpc (both use the same St\"ackel potential adopted in this work), so it is possible that we could still see evidence of the inner-halo population at $r_{apo}$ $\geq$ 35\,kpc. Another possible interpretation is that this [Fe/H] $\sim -1.4$ peak comprises stars accreted from the Sequoia and Gaia-Sausage mergers\textbf{---}this could explain the hint of bimodality seen in this component (two sub-peaks at [Fe/H] $\sim -1.3$ and [Fe/H] $\sim -1.5$)\textbf{---}while the more metal-poor peak represents stars accreted from a series of more minor mergers.

\subsubsection{Prograde vs. Retrograde}\label{subsubsec:pro_retro}

\begin{figure*}
\plottwo{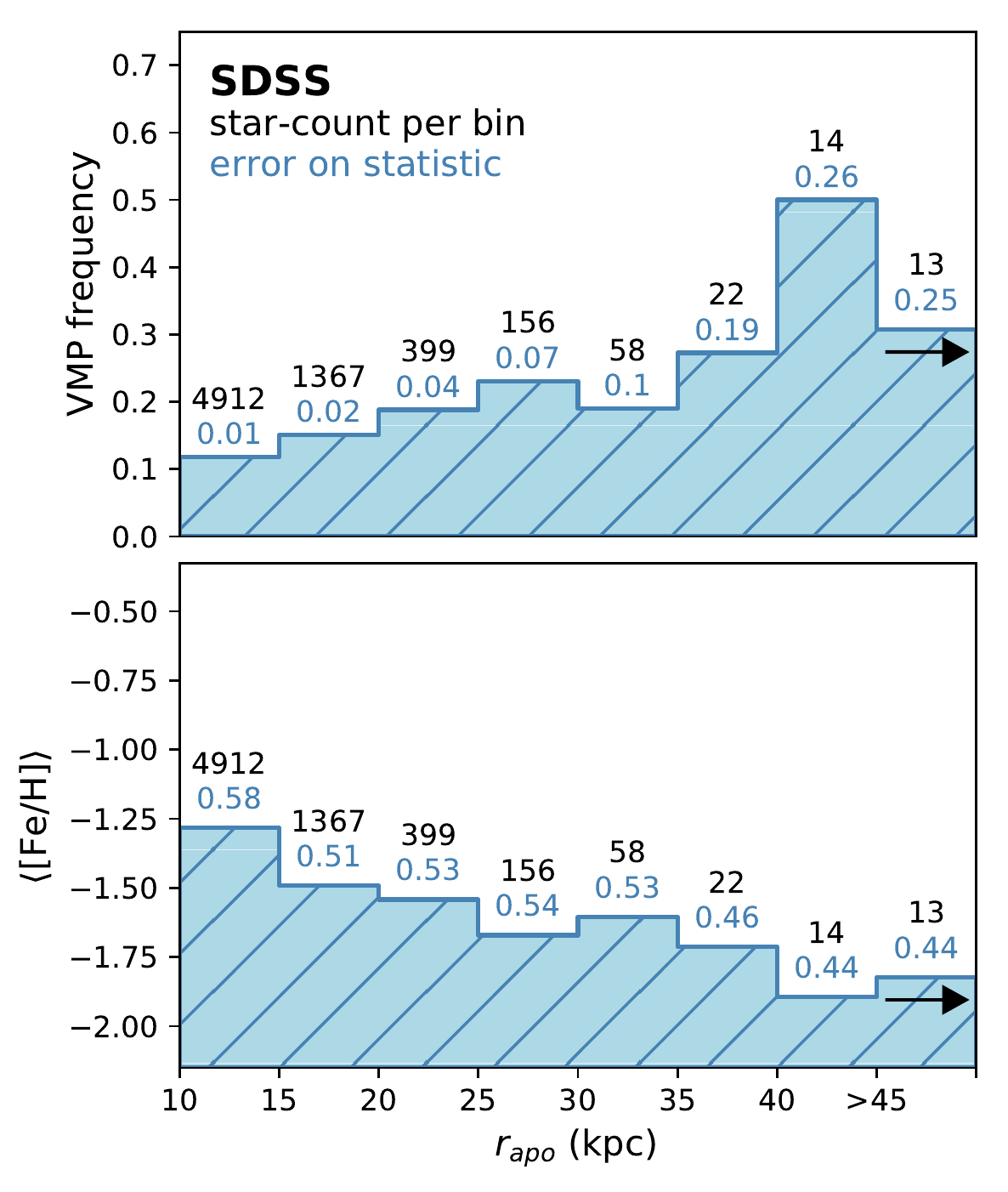}{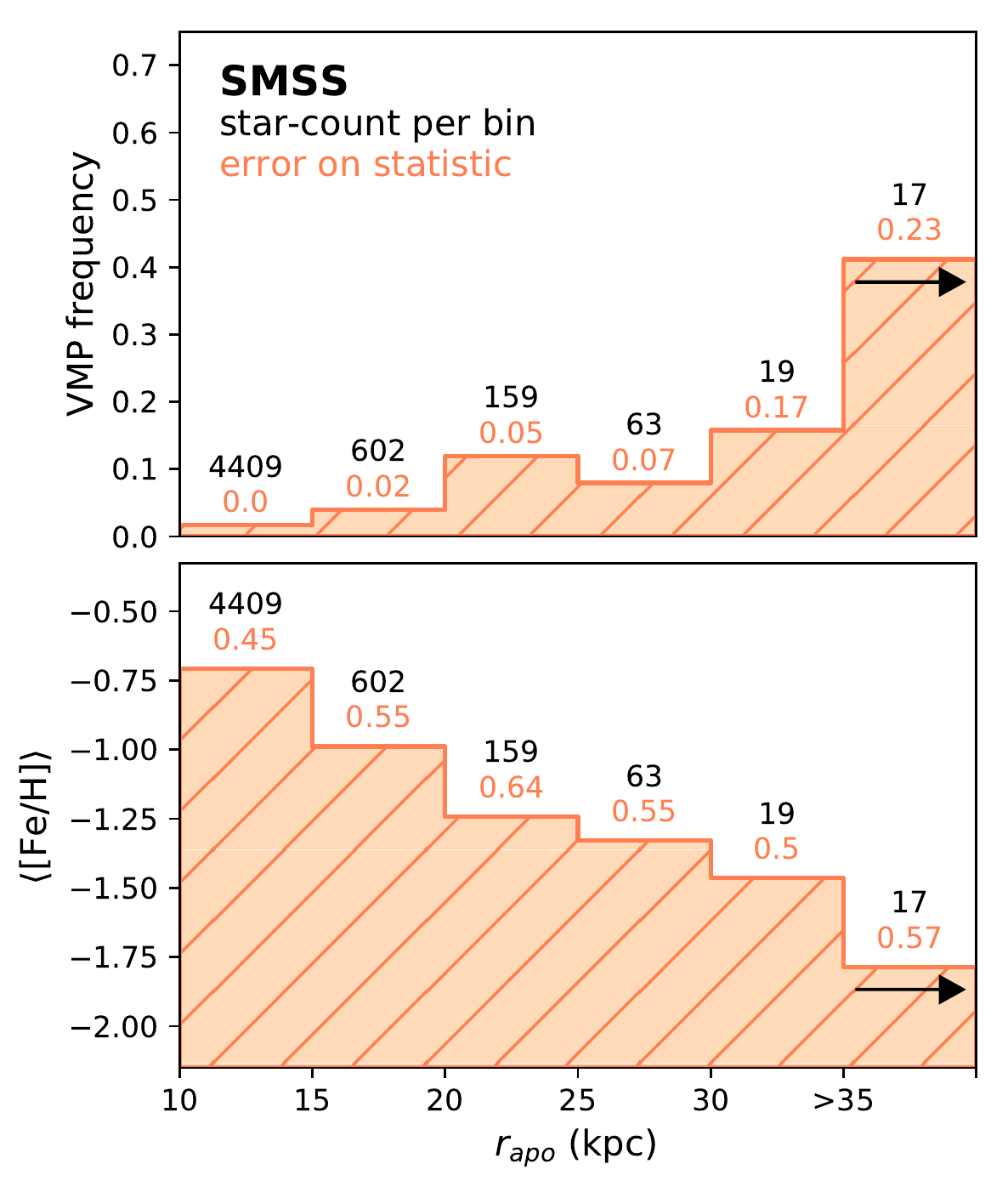}
\begin{center}(a) Prograde sub-samples\end{center}
\plottwo{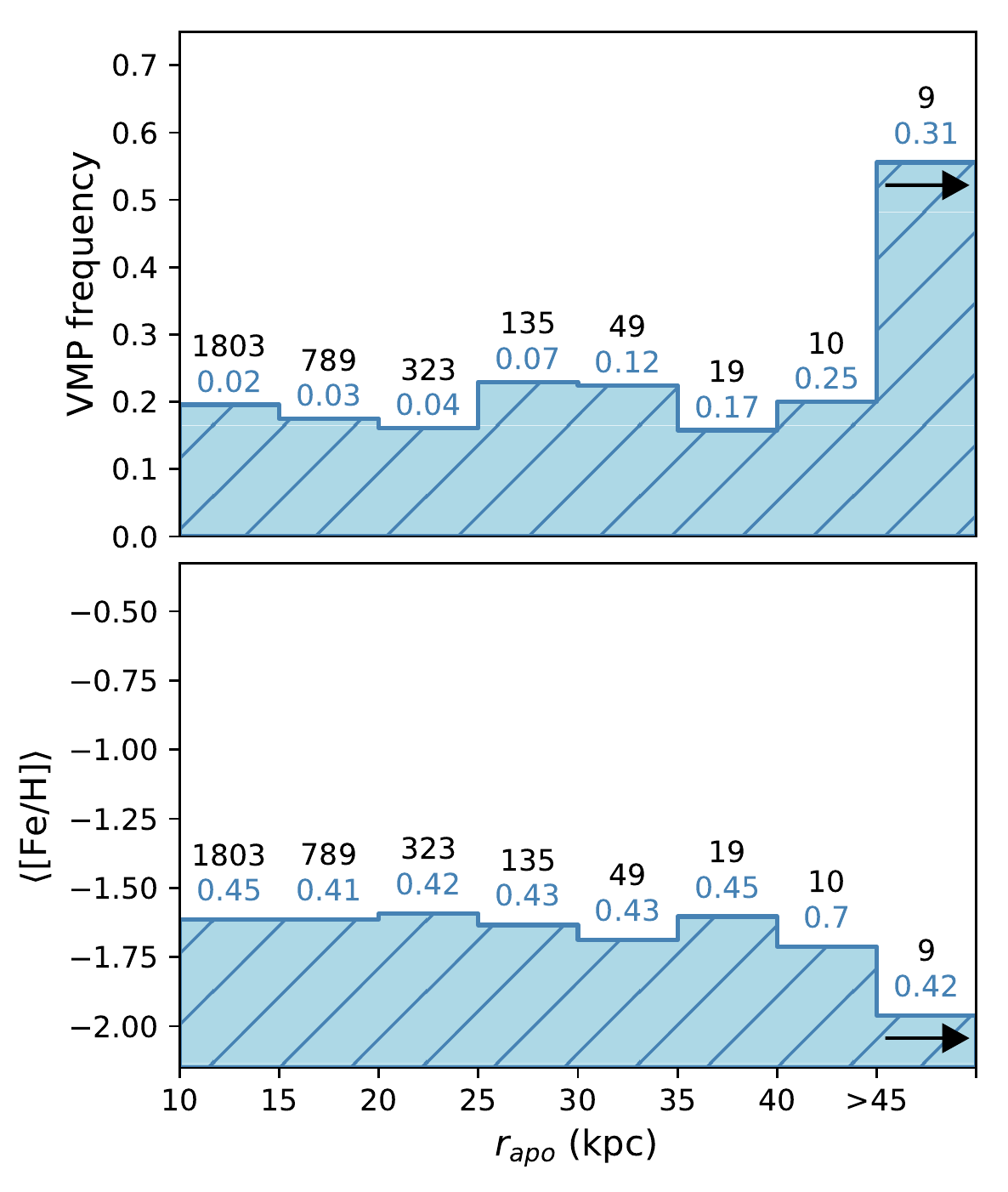}{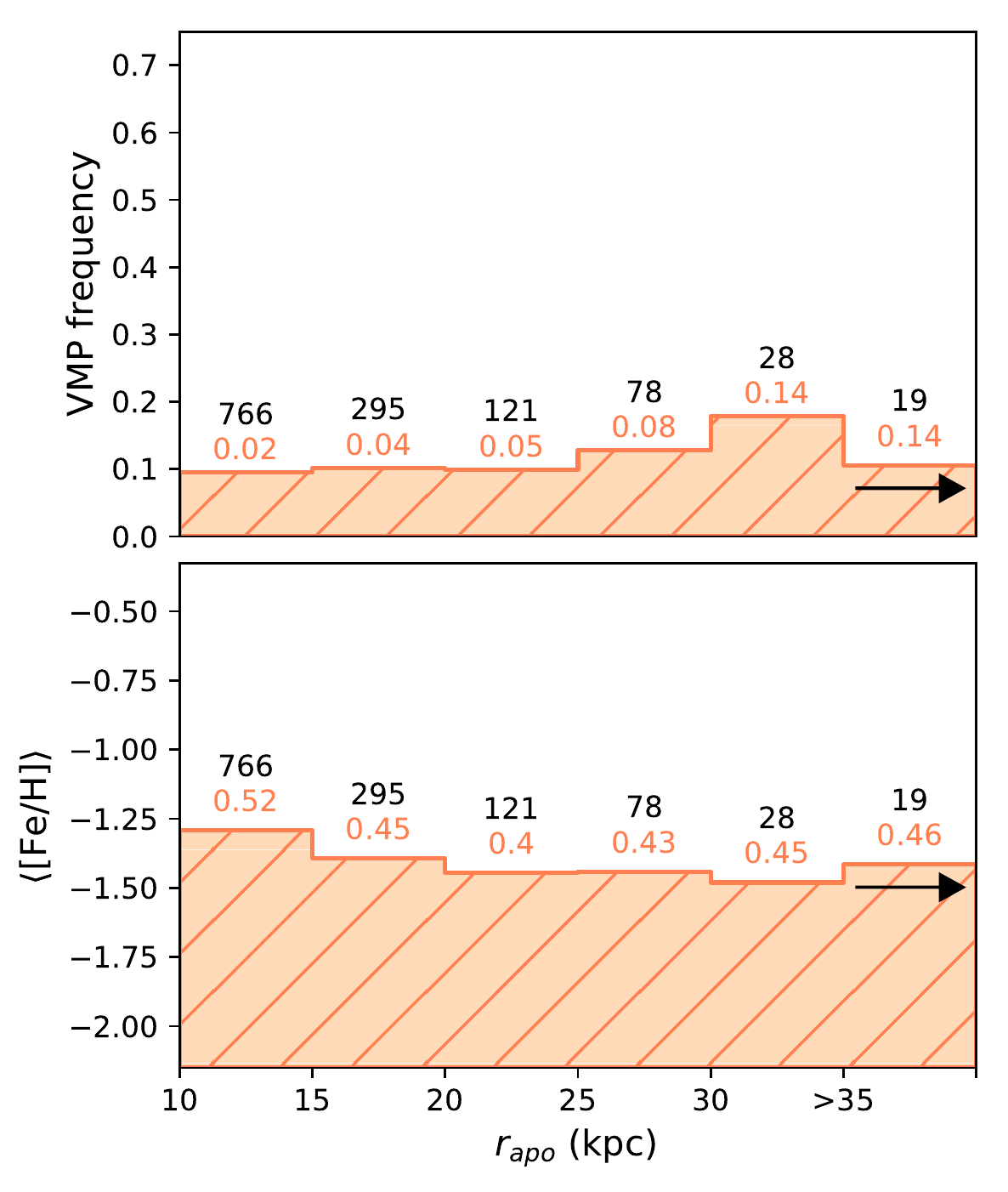}
\begin{center}(b) Retrograde sub-samples\end{center}
\caption{VMP frequencies and average metallicities ($\langle$[Fe/H]$\rangle$) for the SDSS and SMSS samples, divided into (a) prograde and (b) retrograde components. In the VMP frequency panels, star counts and frequency error estimates are indicated for each bin. The error is estimated with a one-sigma Wilson proportion confidence interval \citep{wilson_1927}.  In the mean metallicity panels, star counts and the dispersion of [Fe/H] for each bin are provided.\label{fig:pro_retro}}
\end{figure*}

We can examine these trends in more detail by dividing our local {\it ex-situ} samples on rotational velocity ($v_{\phi}$) into prograde ($v_{\phi}$ $>$ 0\,km\,s$^{-1}$) and retrograde ($v_{\phi}$ $<$ 0\,km\,s$^{-1}$) components, as shown Figure \ref{fig:pro_retro} (a) and \ref{fig:pro_retro} (b), respectively. There are 6,941 prograde stars and 3,137 retrograde stars in the SDSS sample, and 5,267 prograde stars and 1,307 retrograde stars in the SMSS sample. 

The prograde SDSS sub-sample (Figure \ref{fig:pro_retro}(a), left panels) experiences a slow climb in VMP frequency up until $r_{apo}$ $\sim$ 35\,kpc, after which the fraction of VMP stars climbs to $\sim$50\%. Its average metallicity exhibits a steady decrease from 10\,kpc to 45\,kpc. The retrograde sub-sample (Figure \ref{fig:pro_retro}(b), left panels) exhibits no discernible relationship between VMP frequency and $r_{apo}$ in the 10--45\,kpc range, and its average metallicity hovers around $-$1.6 in the same range. Once more, the contrary behavior of the final bins of these sub-samples may be due to low-number statistics (high uncertainty).

The prograde and retrograde SMSS sub-samples (Figure \ref{fig:pro_retro}(a) and (b), right panels) exhibit largely the same behavior as the SDSS sub-samples. The prograde stars show a strong dependence on $r_{apo}$ for both VMP frequency and mean metallicity, while the same quantities for retrograde stars show no noticeable dependence on distance. The Gaia Sausage ([Fe/H] $\sim$ $-1.3$, $L_{z}$ $\sim$ 0\,kpc\,km~s$^{-1}$)  \citep{belokurov_2018,myeong_2018} dominates the retrograde (and prograde) signal in the near halo and the Sequoia progenitor galaxy ([Fe/H] $\sim$ $-1.6$, $L_{z}$ $\sim$ $-2000$ to $-3000$\,kpc\,km~s$^{-1}$ \citep{matsuno_2019,myeong_2019}) likely donated the majority of the retrograde stars in the far halo. The similar peak metallicities and overlapping metallicity ranges (see Figure 2 in \citealt{matsuno_2019} and Figure 9 in \citealt{myeong_2019}) of these imported populations could result in this overall metallicity plateau. 

While the Sequoia (M$_*$ $\sim$ 10$^7$\,M$_\odot$) event may have imprinted a bulk retrograde signal onto the outer-halo, stars donated by numerous small accretion events likely contributed both prograde and retrograde stars to the outer-halo. However, it is likely more difficult to detect (low-metallicity) stars from small mergers in the retrograde outer-halo due to the overwhelming presence of Sequoia stars. In contrast, numerous minor accretions could be the predominant contributors to the prograde outer-halo, based on the metallicity gradient and strongly increasing VMP fraction seen in Figure \ref{fig:pro_retro}. If this behavior accurately reflects the assembly history of the outermost halo, observational efforts to compile catalogs of the most ancient, most metal-poor stars may have more success targeting candidates in the prograde rather than the retrograde outermost halo.

\section{Summary and Conclusions}\label{sec:summ_conc}

We compiled a set of {\it in-situ} ($\sim$21,700 stars in total) and {\it ex-situ} ($\sim$16,500 stars in total) samples to confirm the presence of a metallicity gradient in the outermost halo of our Galaxy and explore the complex assembly history of the Galactic halo. The results of the {\it in-situ} analyses are suggestive regardless of the metallicity-distance selection bias. In our {\it ex-situ} analyses, we find clear evidence of both a negative metallicity gradient over $r_{apo}$ and an increasing relative fraction of VMP stars with distance. In particular, the local SDSS sample exhibits a VMP frequency that reaches $\sim$60\% at 50\,kpc, commensurate with theoretical studies \citep[i.e.,][]{tissera_2014}.

When splitting our samples into prograde and retrograde components, we find that the retrograde appears to exhibit no metallicity-distance correlation while the prograde experiences a steady decline in [Fe/H] and a strong increase in VMP frequency with distance. This may be due to the influence of a more massive merger (metal-richer satellite) in the retrograde direction, versus numerous minor accretions (metal-poorer counterparts) in the prograde direction. As a result, the prograde outermost halo may be the best place to search for the most metal-poor stars.

Since we have placed tight constraints on uncertainty, we may have excluded some stars in the outer-halo that could have given our analysis more significance, but we are also not as likely to detect completely spurious features even at large distances. In addition, our local, {\it ex-situ} samples are not susceptible to the metallicity-distance bias that may affect our  non-local, {\it in-situ} results.

We note here that, recently, \citet{conroy_2019} published an exploration of the Galactic halo using a sample of some $\sim$4,200 giants from the H3 Spectroscopic Survey \citep{H3}, and claimed that no metallicity gradient is detectable in their sample. However, the results of their analyses are not dissimilar from our own findings. Although they find a flat metallicity relation across the majority of the halo, they admit possible evidence for a decreasing mean metallicity beyond $r$ $\sim$ 50\,kpc, which may coincide with the behavior in the $r_{apo}$ $\geq$ 35\,kpc region examined in this work. The VMP component they identify as potentially originating from multiple distinct populations parallels our own hypothesis of a halo component at [Fe/H] $<-2.0$  consisting of numerous accretions of small mini-halos.

Have we fully explored a volume that could qualify as a comprehensive ``outermost" halo, up to the outskirts of the Galaxy? The exact bounds of the outer-halo population are not yet known, and though this work shows the potential for a signature that may extend beyond the volume explored here, further efforts are required to quantify the behavior of the outer-halo MDF beyond 50\,kpc. \textbf{For example, improved kinematics from Gaia DR3 will allow us to expand the narrow magnitude ranges probed by our \textit{ex-situ} samples (see Figure \ref{fig:g_distribs}).} Near-future observations with the Large Synoptic Survey Telescope \citep[LSST;][]{LSST}, combined with spectroscopic follow-up, as well as large spectroscopic surveys undertaken with the Dark Energy Spectroscopic Instrument \citep[DESI;][]{DESI}, the WHT Enhanced Area Velocity Explorer \citep[WEAVE;][]{WEAVE}, and the 4-metre Multi-Object Spectroscopic Telescope \citep[4MOST;][]{4MOST}, will enable a more thorough exploration of the distant halo and the significance of its metallicity gradient.  Further analyses of the VMP catalogs from the LAMOST \citep{LAMOST} survey (e.g., the DR3 VMP catalog of \citealt{li_2018}\footnote{See the ``cleaned" version of this catalog in \citealt{yuan_2019}, as well as the substantially larger DR5 VMP catalog, in preparation.}), should prove illuminating as well.

It would be of particular interest to examine the prograde/retrograde metallicity distributions with a sample of distant carbon-enhanced metal-poor \citep[CEMP;][]{beers_2005} stars. Similarly to metal-poor stars, CEMP stars have been recognized as useful tracers of Galactic formation history \citep[e.g.,][]{beers_2005,carollo_2010,carollo_2014,lee_2017,yoon_2018,lee_2019,yoon_2019}, and their individual \textbf{nucleosynthetic} sub-classes can provide even more information about the origins of various stellar populations. CEMP stars that display over-abundances of elements associated with the slow neutron-capture process (CEMP-$s$) are thought to originate in more-massive galaxies, while CEMP stars that exhibit no over-abundances of neutron-capture products (CEMP-no) are thought to originate in less-massive galaxies \citep[e.g.,][]{lee_2017,yoon_2018,yoon_2019}. A comparison of the CEMP-$s$ to CEMP-no ratios (which can be identified using absolute carbon abundance criteria, readily measured with medium-resolution spectroscopic data, see \citealt{yoon_2016}) in the prograde and retrograde outermost halo could help clarify the origins of the populations present there.

\acknowledgments
We thank \textbf{Xiang Xiang Xue for access to the J16 data-set,} Young Sun Lee for his assistance and discussions regarding SDSS data\textbf{, and the referee for their careful consideration of this work}. The authors acknowledge partial support
from grant PHY 14-30152, Physics Frontier Center/JINA Center for the
Evolution of the Elements (JINA-CEE), awarded by the US National Science
Foundation.

This work has made use of data from the European Space Agency (ESA) mission
{\it Gaia} (\url{https://www.cosmos.esa.int/gaia}), processed by the {\it Gaia}
Data Processing and Analysis Consortium (DPAC,
\url{https://www.cosmos.esa.int/web/gaia/dpac/consortium}). Funding for the DPAC
has been provided by national institutions, in particular the institutions
participating in the {\it Gaia} Multilateral Agreement.

Funding for the SDSS and SDSS-II has been provided by the Alfred P. Sloan Foundation, the Participating Institutions, the National Science Foundation, the U.S. Department of Energy, the National Aeronautics and Space Administration, the Japanese Monbukagakusho, the Max Planck Society, and the Higher Education Funding Council for England. The SDSS Web Site is http://www.sdss.org/.

The SDSS is managed by the Astrophysical Research Consortium for the Participating Institutions. The Participating Institutions are the American Museum of Natural History, Astrophysical Institute Potsdam, University of Basel, University of Cambridge, Case Western Reserve University, University of Chicago, Drexel University, Fermilab, the Institute for Advanced Study, the Japan Participation Group, Johns Hopkins University, the Joint Institute for Nuclear Astrophysics, the Kavli Institute for Particle Astrophysics and Cosmology, the Korean Scientist Group, the Chinese Academy of Sciences (LAMOST), Los Alamos National Laboratory, the Max-Planck-Institute for Astronomy (MPIA), the Max-Planck-Institute for Astrophysics (MPA), New Mexico State University, Ohio State University, University of Pittsburgh, University of Portsmouth, Princeton University, the United States Naval Observatory, and the University of Washington.

Funding for SDSS-III has been provided by the Alfred P. Sloan Foundation, the Participating Institutions, the National Science Foundation, and the U.S. Department of Energy Office of Science. The SDSS-III web site is http://www.sdss3.org/.

SDSS-III is managed by the Astrophysical Research Consortium for the Participating Institutions of the SDSS-III Collaboration including the University of Arizona, the Brazilian Participation Group, Brookhaven National Laboratory, Carnegie Mellon University, University of Florida, the French Participation Group, the German Participation Group, Harvard University, the Instituto de Astrofisica de Canarias, the Michigan State/Notre Dame/JINA Participation Group, Johns Hopkins University, Lawrence Berkeley National Laboratory, Max Planck Institute for Astrophysics, Max Planck Institute for Extraterrestrial Physics, New Mexico State University, New York University, Ohio State University, Pennsylvania State University, University of Portsmouth, Princeton University, the Spanish Participation Group, University of Tokyo, University of Utah, Vanderbilt University, University of Virginia, University of Washington, and Yale University.

\software{astropy \citep{astropy_2013,astropy_2018}, galpy \citep{galpy}, numpy \citep{numpy}, matplotlib \citep{matplotlib}, scipy \citep{scipy}, scikit-learn \citep{sklearn}, STILTS \citep{STILTS}}

\bibliography{dietz_bib.bib}

\end{document}